\documentclass{webofc}
\RequirePackage{orcidlink}

\usepackage[varg]{txfonts} 
\usepackage{caption}
\usepackage{hyperref}
\usepackage{url}

\usepackage{listings}
\hypersetup{colorlinks=true,citecolor=blue,urlcolor=blue,linkcolor=blue}

\begin{document}
\title{IceCube Takes Flight with Pelican --- A First Experience}

\author{\firstname{David} \lastname{Schultz}\inst{1}\orcidlink{0000-0002-7841-3919}\fnsep\thanks{\email{dschultz@icecube.wisc.edu}} \and
        \firstname{Maclean} \lastname{Mansfield-Parisi}\inst{1}\orcidlink{0009-0003-9769-9099}\fnsep\thanks{\email{mmansfie@icecube.wisc.edu}}
}

\institute{Wisconsin IceCube Particle Astrophysics Center, University of Wisconsin--Madison, 222 W Washington Ave, Suite 500, Madison WI 53703, USA   }

\abstract{The IceCube Neutrino Observatory has removed GridFTP and x.509 certificate authentication for data transfers, migrating to the Pelican Platform, the Open Science Data Federation, and WLGC tokens. While this is a common solution on the computing infrastructure we use, we required several customizations to work with our existing data storage structure and make it easier for scientists to use. We wrote a custom WLCG token issuer to support our POSIX filesystem with custom user and group permissions across the entire filesystem. We also made several modifications to HTCondor to ease job submission using tokens, including a custom credmon. After an initially bumpy transition due to several now-resolved Pelican issues, the Pelican-based system has already proven superior to GridFTP in several ways.
}

\maketitle

\section{Introduction}
\label{sec-intro}
The IceCube neutrino detector~\cite{Aartsen:2016nxy} is located at the geographic South Pole and was completed at the end of 2010. It consists of 5160 optical sensors buried between 1450 and 2450 meters below the surface of the South Pole ice sheet and is designed to detect interactions of neutrinos of astrophysical origin. The IceCube Collaboration’s worldwide computing grid consists of many heterogeneous clusters connected into a global pool using HTCondor~\cite{condor2005}, with data transferred back and forth to our main data storage at UW--Madison.

Previously, IceCube data was primarily transferred to and from those heterogeneous clusters via GridFTP~\cite{gridftp, foster2006globus}, with authorization and authentication using X.509 user certificates and proxies~\cite{welch2004x} sent alongside compute jobs. 
GridFTP was phased out beginning in 2019 ~\cite{htc-globus}, with a hard cutoff imposed by the end of CILogon~\cite{Basney:2019Rj} X.509 certificate issuance in May 2025. We were not informed about the cutoff until after it had occurred, and the inability to renew existing tokens expedited our transition.

\subsection{Pilot Projects}
\label{subsec-pilots}

We first started investigating new technologies in 2019, when the Open Science Grid (OSG)~\cite{osg07} informed operators about plans to end support for GridFTP~\cite{htc-globus}. At that time, no hard cutoff for access was determined, only lack of ongoing support from May 2022 onward, so transition was not a priority. We discussed with community partners and agreed that HTTP-based transfers using OAuth2~\cite{rfc6749} tokens were likely the way forward. Progress was immediately stymied by the COVID-19 pandemic in 2020 and focus on the project was lost.

We resumed this project in 2022 with an attempt to use Nginx's~\cite{reese2008nginx} built-in WebDAV~\cite{rfc4918} support. Using one of the world's most widely used web servers has significant advantages in security and maintainability, but the authentication and authorization was difficult. Our single sign-on system for the IceCube Collaboration uses Keycloak, and we were able to configure it to generate OAuth2 tokens containing the requesting user's UID and GIDs. Then using a Lua script, Nginx could switch to that UID to attempt to read from or write to files. While it worked, it was fragile and continued to use the user proxy method from X.509 but in token format --- embedding identity rather than capability.

In 2024 a new project was developed to directly support WLCG-style OAuth2 tokens~\cite{wlcg-jwt}, which use scopes to define authorization capabilities. This would allow for more interoperability in the community, such as supporting dCache~\cite{fuhrmann2006dcache,millar2012dcache}. We did some initial evaluation using Keycloak to create these tokens with hard-coded scopes for a specific user, connecting to a dCache instance at DESY-Zeuthen. While the tests were successful, the scope creation was limited.

A deadline materialized in 2025 when CILogon (our source of certificates supported by GridFTP) stopped signing new X.509 certificates, and its absence we chose to use the Pelican Platform~\cite{pelican} and the Open Science Data Federation (OSDF)~\cite{andrijauskas2026new}. This was already integrated into HTCondor and the Open Science Grid (OSG), which we use heavily. It also provided a command-line tool for users, with built-in OAuth2 identity provider lookup. As such, though Pelican was a fairly new project, we believed it would work without too much difficulty.

\subsection{Keycloak as a Token Issuer}
\label{subsec-keycloak}

Since we already had experience generating WLCG-style OAuth2 tokens with Keycloak, we first decided to try to extend that systen to the dynamic scopes that are necessary for Pelican transfers. Keycloak does have some support for this, but required us to write a custom plugin to verify whether a user is authorized for specific scopes. We wrote this plugin using their Javascript connector, and it seemed to work in initial testing. Unfortunately, we found a critical bug in Keycloak's dynamic scope handling, where it will only allow a single dynamic scope assignment per user, across all active sessions. If a new scope is requested, it removes the old scope, and existing refresh tokens will be generated with the new scope on the next refresh. This was a fatal flaw for us, which caused us to abandon Keycloak as a token issuer for Pelican, although we continue to use it as an identity provider (IDP) in other systems and for interactive Pelican requests.

\section{Data Transfer Architecture}
\label{sec-arch}

\begin{figure}[h]
\centering
\begin{minipage}[b]{.48\textwidth}
  \centering
  \includegraphics[width=.7\linewidth]{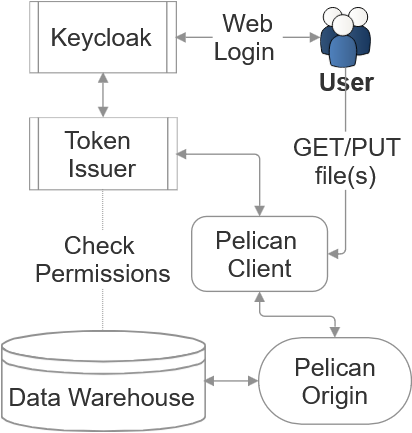}
  \captionof{figure}{Pelican command-line client flow. A user requests to get or put a file, and the Pelican client first contacts the token issuer to perform device authorization. The user is prompted to login with our Keycloak IDP in the browser, and the requested scopes are checked by the issuer against POSIX permissions. The Pelican client will then receive refresh and access tokens and use the access token as authorization with the Pelican Origin, which will check that it is valid before getting or putting the file.}
  \label{fig-1}
\end{minipage}%
\hfill
\begin{minipage}[b]{.48\textwidth}
  \centering
  \includegraphics[width=.9\linewidth]{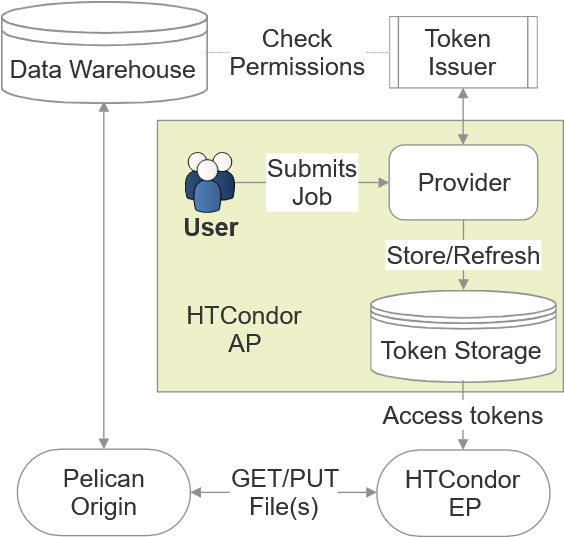}
  \captionof{figure}{HTCondor token flow. The user submits an compute job, which uses our custom HTCondor token provider to request a token on their behalf. Because our provider is a privileged client, it impersonates the user to get refresh and access tokens for them. Our provider has a separate process (HTCondor's CredMon) to refresh these tokens periodically while jobs are still in the queue. The access token is then sent with the job to the execution point to use for file transfers with the built-in Pelican client plugin.}
  \label{fig-2}
\end{minipage}
\end{figure}

A quick summary of the data transfer architecture is warranted. First, we have a token issuer, which we chose to design ourselves, as described in Section \ref{subsec-issuer}. Then, there are the data transfer servers, a set of Pelican Origins in front of our POSIX filesystem. In the case of a user interactively accessing data through Pelican, the Pelican client can GET files directly from our Origins or via caches, and PUT files to our Origins. More details on the Pelican Origins can be found in \ref{subsec-pelican}, and the command-line client is diagrammed in Figure \ref{fig-1}. For batch jobs, HTCondor manages tokens on behalf of the user, requesting them, refreshing them, sending access tokens with jobs, and triggering the Pelican client to transfer files in and out. There are several components to this, which are covered in Section \ref{sec-condor} and diagrammed in Figure \ref{fig-2}.

\clearpage

\subsection{IceCube Token Issuer}
\label{subsec-issuer}

As a result of our previous projects, we designed and deployed a OAuth2 token issuer tailored for IceCube storage workflows. We had a few requirements: 

\begin{enumerate}
    \setlength{\topsep}{0pt} 
    \setlength{\itemsep}{0pt}
    \item Must delegate identity verification upstream to our Keycloak identity provider for single sign-on capabilities.
    \item Must validate requested WLCG scopes against our POSIX filesystem user/group permissions.
    \item Must support OAuth2 device authorization~\cite{rfc8628} for the Pelican command-line client.
    \item Must support OAuth2 dynamic client registration~\cite{rfc7591} for the Pelican command-line client.
    \item Must support OAuth2 token exchange~\cite{rfc8693} to:
    \begin{enumerate}
        \setlength{\itemsep}{0pt}
        \item Allow other clients to give HTCondor a refresh token.
        \item Restrict the scope of tokens to send the minimal scopes necessary in an access token to a job.
        \item Allow privileged clients to perform impersonation and generate tokens on behalf of users in special cases.
    \end{enumerate}
\end{enumerate}

IceCube already had a robust OAuth2 client library with utilities for generating and validating JWTs in Python. This was built upon, with a new Python server to handle the OAuth2 flows and a library for generating and rotating signing keys. The issuer server also includes a library to validate requested scopes and users against the POSIX filesystem, which uses a TTL cache for LDAP lookups of user and group details.

For the OAuth2 Authorization Code and Device Authorization flows, the user must use a browser to authenticate with the token issuer. As part of this process, they are automatically forwarded to our Keycloak IDP server to log in. This keeps the token issuer minimal in terms of user interaction, only providing tokens instead of managing user logins.

To handle dynamic clients and refresh tokens, state is stored in MongoDB~\cite{chodorow2010mongodb}. This data is only temporary, and is cleared out periodically after it has expired. Permanent clients, and clients that are allowed to perform impersonation, are registered via environment variables at application startup. User impersonation via token exchange is handled via an extension to the OAuth2 RFC by Keycloak~\cite{keycloak-impersonation}.

Due to the sensitive nature of the token issuer, we've thoroughly tested its software, with extensive code coverage. We use automatic code scanning on every commit and periodically evaluate the code with the Software Assurance Marketplace~\cite{kupsch2017continuous,swamp}, using modern LLM scanning tools designed specifically to find security issues. We are confident in the robustness of the final product.

\subsection{Data Serving Architecture}
\label{subsec-pelican}

The Pelican Origin is a server daemon which serves requests to get or put data over HTTPS. It communicates with other parts of a Pelican federation to establish its location for a given data path, both physical and on the network, and to cache data on distributed cache servers. In registering with a federation, a namespace is assigned that an Origin can then serve, in our case \texttt{osdf:///icecube/wipac}. The Pelican Origin data transfer is agnostic of the storage backend, having options to serve and store data in POSIX filesystems, S3 object storage, and other options. We use CephFS as a POSIX network filesystem as our backing storage, shared with other internal systems. In order to access data restricted to POSIX users and groups, the Pelican Origin is able to switch users and access the filesystem as any user. We use LDAP for shared identity across our servers and filesystems, and Pelican is able to map the identity field in each token to an LDAP username. 

We federate our data using the Open Science Data Federation (OSDF), a Pelican federation operated by the Open Science Grid (OSG). OSDF caches are spread geographically and often close to grid computing sites we commonly use. Irrespective of GridFTP deprecation driving us to adopt Pelican, its caching and checksum validation features have been useful to us on grid computing systems to improve data locality and reliability.

\vspace{10pt}

\section{HTCondor Integration}
\label{sec-condor}

HTCondor includes some built-in mechanisms to handle OAuth2 tokens and data transfer. HTCondor provides a native file transfer protocol between the Access Point (AP) job submission server and the Execution Point (EP) where the job runs, and the mechanisms for managing these transfers have been extended to perform transfers using external services. URLs can be specified for transfers, with the scheme specifying which plugin should be used. As of recent HTCondor releases, Pelican is included as a built-in plugin type. With a built-in plugin, allowing users to perform transfers is a matter of configuration and token handling.

There are several key terms in the transfer configuration. A Provider is a named source for token configuration, including the OAuth2 issuer url, client id, and secret. A Handle is an optional name for a specific token with scopes. If not specified, a user only has one token. A CredMon is the daemon that handles token refresh.

If our Provider is named \texttt{pelican} and we use a Handle named \texttt{myjob}, an HTCondor submit file will have lines like:

\vspace{-3pt}
\begin{lstlisting}
use_oauth_services = pelican
pelican_oauth_permissions_myjob = storage.read:/data/exp/IceCube
transfer_input_files = pelican.myjob+osdf:///icecube/wipac/data/...
\end{lstlisting}
\vspace{-3pt}

This will call our Provider and create a custom token named \texttt{myjob} with the specified storage read scope. Our Provider runs as an impersonation client and uses the current username of the job submitter. We have chosen this because every user submitting to our HTCondor AP has already authenticated by logging in to the machine, and we trust that login as it is tied directly to our IDP. In the event a user is not authorized for a specific scope the token creation will fail, causing the job submission to fail as well. This gives the user an error message as soon as possible, and prevents jobs with unauthorized scopes to run and then fail during the file transfer.

\subsection{Job Transforms and User Abstraction}
\label{subsec-job-transforms}

HTCondor has several mechanisms for an Access Point administrator to define transforms to run on every job submission. We use these to help abstract some of the more cumbersome syntax details from users, although they may specify them directly at any time. In particular, we have one template that helps define the Provider, Handle, and token scopes. Another transform can convert from our POSIX filesystem to valid URLs and transfer syntax.

\vspace{-3pt}
\begin{lstlisting}[caption={User's compact job submission}, label={lst:compact}]
use template : Oauth(myjob, /data/exp/IceCube, /data/user/dschultz)
transfer_input_files = /data/exp/IceCube/foo.txt
transfer_output_files = /data/user/dschultz/bar.txt
\end{lstlisting}
\vspace{-3pt}

\vspace{-3pt}
\begin{lstlisting}[caption={Expanded HTCondor job submission}, label={lst:expanded}]
use_oauth_services = pelican
pelican_oauth_permissions_myjob = storage.read:/data/exp/IceCube #\\#storage.write:/data/user/dschultz
requirements = stringListMember("pelican", HasFileTransferPluginMethods)
transfer_input_files = #\\#pelican.myjob+osdf:///icecube/wipac/data/exp/IceCube/foo.txt
transfer_output_files = bar.txt
transfer_output_remaps = #\\#bar.txt = pelican.myjob+osdf:///icecube/wipac/data/user/dschultz/bar.txt
\end{lstlisting}

\subsection{Custom Provider and CredMon}
\label{subsubsec-rust}

We tried to make the built-in HTCondor OAuth2 provider and CredMon work for us, but failed to get it to work with Handles and token exchange. We were suggested by a member of the HTCondor team to write these ourselves and we did, writing these in Rust as statically compiled binaries.

We picked Rust for a few reasons. First, Rust's memory safety gave us security in the multi-user and unpredictable environment of grid computing. They also had to execute quickly, especially as the Provider runs on every job submission. Being statically compiled, binaries will work across Linux distributions without an issue. And finally, static binaries allow us execute the Provider with setuid, allowing it to run as root and read the protected client secret file.

When the Provider is called, it loads the OAuth2 client id and secret from the HTCondor server configuration and gets the username of the user submitting the job. First it checks on disk in the HTCondor credential storage if this token already exists with the requested scopes. If not, it requests new tokens from the token issuer. If this request fails, it sends the error message to the user and fail the job. Otherwise it stores the received tokens in HTCondor credential storage.

For periodic token refresh, another custom process called the CredMon runs as part of the HTCondor daemon set. It can be configured to periodically check the expiration of all access tokens and refresh them if necessary. Refreshed access tokens for actively running jobs will automatically be sent to the Execution Point at an interval (by default, 240 seconds). This is actually a hard limit on the minimum token validity window, as access tokens valid less than this interval may expire on the EP before being updated and fail a transfer. In practice, the minimum access token expiration time should be larger than the sum of both the CredMon interval and EP update interval. While 300 seconds is the default minimum, 600 seconds has proven more practical in case of transient errors.

\section{Production Issues \& Lessons Learned}
\label{sec-prod}

While this system generally worked well from the beginning, there were a few teething issues. One version of the Pelican client had improper error handling on the command-line, returning with an exit code of 0 (success) even when it printed a failure. The Pelican team was able to fix this in a single release once notified. There were also breaking changes released in how Pelican handled overwriting files, disallowing them on the client. Due to caching, overwriting files in a Pelican federation has been a complicated issue. Because HTCondor jobs that restart during a file transfer may need to overwrite a partially written file, we convinced the Pelican team to add an option to allow overwriting. Other minor issues included adding load balancing between multiple Pelican Origin servers and an integer overflow bug with transfer IDs in long-running Origins.

Some operational challenges we discovered while operating the system for several months were caused by a dependency on the OSDF Pelican Director. This federation server must be available or all requests will fail. While outages have been frequent, this was a change from our previous GridFTP setup which had no external network dependencies, and was something for our operational team and users to get used to. The OSDF also employs substantial read caching, which can occasionally produce errors.

Errors reported back to users when transfers fail could sometimes be rather obscure. Due to Pelican retry logic, there could be multiple errors for a single transfer, sometimes returning tens of lines across both a user message and an admin log message. We've tried highlighting specific error text as examples in our user documentation, to give users a guidance on whether errors are a permission error, a badly requested scope, or a transient error they should ignore and resubmit the job for. We've also had to teach users that certain special characters (e.g., colons or underscores) cannot be used in HTCondor token Handle names, which should ideally be an ASCII alphanumeric string that is fairly short.

\section{Future Work}
\label{sec-future}

While we have a working system now, there are still several areas of future work. We're working with the HTCondor and Pelican teams on ways to make clearer error messages when things go wrong. Both the command-line client and HTCondor components sometimes lack simple, plain language that a user can understand and act on, and often lack indications on what they should do to resolve the error. For some errors, the solutions are obvious enough on a technical level that we would like to automate their fixes, especially in the case of an Execution Point error where the job only needs to retry a transfer in another way rather than failing the completed job.

Submission validation is another focus area. If we can catch submission errors in requested transfers early, before running a job, we can save computing time and lower the latency before the error is raised to the user. There is also potential to improve the user's experience letting them specify files more naturally before converting those definitions into the ClassAds HTCondor needs for data transfer.

\section{Conclusion}
\label{sec-conclusion}

The migration of the IceCube grid computing data transfers from legacy GridFTP and X.509 certificates to the Pelican Platform, Open Science Data Federation (OSDF), and WLCG tokens has been largely successful and is working well. With a custom token issuer, custom HTCondor components and suitable job submission transforms, we were able to maintain reliable and performant data transfers to grid computing sites. The transition surfaced several lessons in external service availability, error verbosity, and token handling, but the resulting system has already demonstrated significant usability improvements over previous data transfer infrastructure.

\section*{Acknowledgments}

This work was partially funded by the U.S. National Science Foundation (NSF) under grant OPP-2042807.

\bibliography{bib}

\end{document}